\documentclass{JHEP3}
\usepackage{epsfig,multicol,bbm,latexsym}
\usepackage{graphicx}

\title{Search for dark matter signals with {\it Fermi}-LAT
observation of globular clusters NGC 6388 and M 15}

\author{Lei Feng$^{a,b,c}$, Qiang Yuan$^a$, Peng-Fei Yin$^a$,
Xiao-Jun Bi$^a$ and Mingzhe Li$^{b,c}$ \\
$^a$Key laboratory of particle astrophysics, Institute of High Energy
Physics, Chinese Academy of Sciences, Beijing 100049, China\\
$^b$Department of Physics, Nanjing University, Nanjing 210093, China\\
$^c$Joint Center for Particle, Nuclear Physics and Cosmology, Nanjing
University -- Purple Mountain Observatory, Nanjing 210093, China\\

fenglei@chenwang.nju.edu.cn, yuanq@ihep.ac.cn, yinpf@ihep.ac.cn,
bixj@ihep.ac.cn, limz@nju.edu.cn
}

\abstract{
The globular clusters are probably good targets for dark matter (DM)
searches in $\gamma$-rays due to the possible adiabatic contraction of
DM by baryons. In this work we analyse the three-year data collected by
{\it Fermi} Large Area Telescope of globular clusters NGC 6388 and M 15
to search for possible DM signals. For NGC 6388 the detection of
$\gamma$-ray emission was reported by {\it Fermi} collaboration, which
is consistent with the emission of a population of millisecond pulsars.
The spectral shape of NGC 6388 is also shown to be consistent with a
DM contribution if assuming the annihilation final state is $b\bar{b}$.
No significant $\gamma$-ray emission from M 15 is observed. We give
the upper limits of DM contribution to $\gamma$-ray emission in both
NGC 6388 and M 15, for annihilation final states $b\bar{b}$, $W^+W^-$,
$\mu^+\mu^-$, $\tau^+\tau^-$ and monochromatic line.
The constraints are stronger than that derived
from observation of dwarf galaxies by {\it Fermi}.}


\preprint{1112.2438}

\begin{document}

\section{Introduction}

A standard model of cosmology is developed, in which the universe
consists of $4\%$ ordinary baryonic matter, $\sim 23\%$ dark matter
(DM), $\sim 73\%$ dark energy, and a tiny abundance of relic neutrinos
\cite{2011ApJS..192...18K}. The nature of DM particle remains a mystery.
One of the leading candidates is the weakly interacting massive particle
(WIMP), which is predicted in several models, such as neutralino in
supersymmetry model (see the reviews \cite{1988ARNPS..38..751P,
1996PhR...267..195J,2005PhR...405..279B}). In this kind of models, the
mass and interaction strength of DM particles can produce the
correct relic density of DM if the WIMPs are thermally ``freeze-out'',
which is called ``WIMP miracle''.

If DM particles annihilate or decay into standard model particles,
they can be detected indirectly from the cosmic ray (CR) radiation.
Among many kinds of CR particles, $\gamma$-rays are the best probe
due to their simple propagation. {\it Fermi} gamma-ray telescope,
which was launched in 2008, has surveyed the $\gamma$-ray sky with
very high resolution and sensitivity for more than three years.
Nearly $2000$ sources as well as the diffuse $\gamma$-ray emission
were detected by {\it Fermi} Large Area Telescope ({\it Fermi}-LAT)
\cite{2011arXiv1108.1435T,2009ApJ...703.1249A,2009PhRvL.103y1101A,
2010PhRvL.104j1101A}. The analysis of the {\it Fermi}-LAT data in
the Galactic center region did see some excesses with respect to
the background model \cite{2010ApJ...717..825D,2010ApJ...724.1044S},
however, there is no strong indication of signals from DM annihilation
or decay\footnote{See also the argument of possible DM explanation
of the $\gamma$-ray haze/bubble \cite{2011ApJ...741...25D}.}.
The constraints on DM model parameters can be derived according to
the non-detection of DM signals from e.g., dwarf galaxies
\cite{2010ApJ...712..147A,2010PhRvD..82l3503E,2011PhRvL.107x1303G,
2011PhRvL.107x1302A,2011arXiv1111.2604C}, galaxy clusters
\cite{2010JCAP...05..025A,2010PhRvD..82b3506Y,2010JCAP...12..015D,
2011PhLB..698...44K,2011arXiv1105.3240P,2011arXiv1110.1529H} and the
diffuse $\gamma$-rays
\cite{2010JCAP...04..014A,2010NuPhB.840..284C,2010JCAP...03..014P,
2010JCAP...06..027Z,2010JCAP...07..008H,2010JCAP...10..023Y,
2010JCAP...11..041A,2010arXiv1011.5090A,2011arXiv1105.4230C,
2011PhRvD..83l3513Z,2011JCAP...04..020Y}.

Due to the very weak interactions of DM particles, it is important
to investigate the sites with high DM density when searching for DM
annihilation signals. The proposed good candidates include the Galactic
center, dwarf galaxies, Galactic subhalos and cluster of galaxies.
The Milky Way globular clusters (GCs), defined as spherical
ensemble of stars that orbits the Galaxy as satellites, are also
potentially good targets for indirect detection of DM. The formation
of GCs remains a poorly understood problem. There are generally two
scenarios to describe the formation of GCs. The primordial formation
scenario suggests that GCs were formed in cosmological DM minihalos
before the formation of galaxies \cite{1984ApJ...277..470P}. The other
way to form GCs might be the star-forming events such as the merger of
galaxies. There was evidence to show that metal-poor GCs might have
a cosmological origin and metal-rich GCs might form in the galaxies
\cite{2006ARA&A..44..193B}. If the GCs were formed in cosmological
DM minihalos, they would experience the adiabatic contraction (AC) due
to the infall of baryons during the evolution of GCs and leave a high
density spike of DM. GCs are not usually discussed for DM detection
due to the poor knowledge about their origin and the observational fact
that there is in general no significant amount of DM in vicinity of GCs
\cite{2009MNRAS.396.2051B,2010MNRAS.406.2732L,2010arXiv1010.5783C}.
However, there is possibility that the high density spike of DM due to the
AC process may still play an important role for the annihilation signals.
The previous works to search for or constrain DM models with $\gamma$-rays
from GCs include \cite{2008PhRvD..78b7301Z,2008ApJ...678..594W}.

Recently the atmospheric Cherenkov telescope array High Energy Stereoscopic
System (H.E.S.S.) had investigated two GCs NGC 6388 and M 15 to search
for possible DM signals \cite{2011ApJ...735...12A}.
No $\gamma$-ray signal was detected by H.E.S.S. and strong constraints on
the DM model parameters were given. In this work, we use the three-year data
of {\it Fermi}-LAT to study the $\gamma$-ray emission from DM annihilation
in these two GCs. Detections of $\gamma$-ray emission from some GCs with
{\it Fermi}-LAT were reported \cite{2009Sci...325..845A,2010A&A...524A..75A,
2010ApJ...712L..36K,2011ApJ...729...90T}, including NGC 6388 studied here.
For M 15 there is no detection yet. In this work we will focus on the
possible DM component of the $\gamma$-ray emission, if any, from the GCs.
The upper limits of DM contribution will be derived and the constraints
on DM model parameters will be presented.

\section{Gamma-rays from DM annihilation in globular clusters}

M 15 is a metal-poor GC which favors a cosmological origin of it
\cite{1996AJ....112.1487H}. For NGC 6388, there is strong evidence to
show the existence of an intermediate mass black hole (IMBH) with mass
$\sim 6\times10^3$ M$_{\odot}$ \cite{2007ApJ...668L.139L}, which also
suggests a cosmological origin even though the metallicity is relatively
high \cite{1996AJ....112.1487H}. Therefore we have good motivation to
search for the possible DM annihilation signal from these two GCs. The
estimated stellar masses of NGC 6388 and M 15 are $10^6$ and $5\times10^5$
M$_{\odot}$, with distances $11.5$ and $10.0$ kpc respectively
\cite{2007ApJ...668L.139L,2008ApJ...678..594W}. Other parameters of 
them can be found in Table 2 of Ref. \cite{2011ApJ...735...12A}.

\subsection{DM density distribution}

For the purpose of this work, these two GCs are assumed to form in the
cosmological context, which were DM dominated in the primordial stage,
before reionization and the galaxy formation \cite{1984ApJ...277..470P}.
The AC process of baryons to form the GC is expected to pull DM into
the center and results in a high density core of DM
\cite{1986ApJ...301...27B}. After the AC process the heating effect of
DM due to scattering with baryons will tend to sweep out the high density
DM core, leaving a constant density \cite{2008PhRvD..77d3515B}.
The IMBH, if exists, may further modify the density profile through
adiabatic accretion \cite{2005PhRvL..95a1301Z}.

The modelling of the GC DM halo can be divided into three steps. The
first step is the AC process of the dark halo during the collapse of the 
core of GC. Supposing that the DM particles travel on circular orbits, 
the enclose mass distribution of DM $M(r)$ can be calculated with the 
follow equation \cite{1986ApJ...301...27B}
\begin{equation}
[M_{{\rm DM},i}(r_i)+ M_{b,i}(r_i)]r_i = [M_{{\rm
DM},f}(r_f)+M_{b,f}(r_f)]r_f,
\end{equation}
where the subscript $i(f)$ denotes the initial (final) mass distribution 
of baryon or DM. The initial mass of the minihalo is assumed to be $10^7$ 
M$_{\odot}$, with Navarro-Frenk-White (NFW, \cite{1997ApJ...490..493N}) 
density profile for both the DM and baryon distributions\footnote{Note 
that for such minihalos the density profile might be smoother
\cite{2009MNRAS.397.1169D,2011arXiv1111.1165S}, however, as shown in
\cite{2008PhRvL.100e1101S} the initial density profile does not
affect significantly the final DM profile after AC process.}. The
mass fraction of baryons is adopted to be $20\%$. For the
convenience of comparison, these adoptions are the same as that in
Ref. \cite{2011ApJ...735...12A}. We should keep in mind that these
parameters may have large uncertainties and the quantitative results
of this work may also suffer from uncertainties.

Given the final baryon distribution, which can be derived according to
the observed surface density distribution of the GC\footnote{See, e.g.,
http://www.physics.mcmaster.ca/~harris/mwgc.dat}, one can get the DM
density profile after AC \cite{1986ApJ...301...27B}. The final baryon
density for NGC 6388 is taken from \cite{2011ApJ...735...12A}, which
was computed using the surface density profile given in
\cite{2007ApJ...668L.139L}. For M 15 the final baryon density is
taken from \cite{1997AJ....113.1026G}.

The second step is to take into account the smoothing effect due to
baryon heating after AC process. For the convenience of discussion, we 
employ the relaxation time $T_r$ defined as \cite{1987degc.book.....S}
\begin{equation}
T_{r} = \frac{3.4 \times 10^{9}}{ \ln{\Lambda}}\left(\frac{v_{\rm rms}}
{\rm km\,s^{-1}}\right)^3 \left(\frac{m}{\rm M_{\odot}}\right)^{-2}
\left(\frac{n}{\rm pc^{-3}}\right)^{-1}{\rm yr} \; ,
\label{eq:relaxationtime}
\end{equation}
where $v_{\rm rms}$ is the velocity dispersion of stars, $m$ is the
typical stellar mass in the GC, $n$ is the stellar number density, and
$\ln{\Lambda}$ is the Coulomb logarithm. $T_r$ is estimated to be
$\sim 7\times 10^4$ yr in the central region of M 15, and $\sim 8\times
10^6$ yr for central NGC 6388 \cite{2011ApJ...735...12A}. The relaxation
time is an increasing function of the distance to the center.
The DM will be heated up due to the scattering with the stars, which
will lead to the dissipation of the DM core \cite{2004PhRvL..92t1304M}.
The scattering time scale is comparable to $T_r$. The heating radius,
$r_{\rm heat}$, is then defined with $T_r(r_{\rm heat})=t_{\rm age}$,
where $t_{\rm age}$ is the age of the Universe. Therefore at small
radius $r<r_{\rm heat}$, the relaxation time is shorter than the age of
the Universe and the heating effect on DM is important. At large radius
$r>r_{\rm heat}$ the DM distribution is unaffected by heating. The
heating radii are estimated to be about $5$ pc and $4$ pc for M 15 and
NGC 6388 respectively. Roughly speaking we have the DM density distribution
with baryonic heating
\begin{equation}
\rho(r)=\left\{
\begin{array}{ll}
\rho_0, & r<r_{\rm heat},\\
\rho_0\times\frac{\rho_{\rm AC}(r)}{\rho_{\rm AC}(r_{\rm heat})},
& r>r_{\rm heat},
\end{array}\right.
\end{equation}
where $\rho_{\rm AC}(r)$ is the DM density profile after AC, which can be
solved with Eq. (2.1), and $\rho_0$ is the density at $r=r_{\rm heat}$.

The third step is to consider the effect of the IMBH, if exists. 
The AC profile of DM will not be significantly affected by the IMBH due
to its small mass compared with the total baryon mass. However, the
following adiabatic accetion of IMBH will modify the DM density profile
after dynamic heating. The radius within which the IMBH is gravitational
dominant is defined by $M(< r_h)=\int^{r_h}_0\rho(r){\rm d}^3r =
2M_{\rm IMBH}$. Then the DM distribution can be expressed in three regions
\begin{equation}
\rho(r)=\left\{
\begin{array}{ll}
\rho_0(r/r_h)^{-3/2}, & r<r_h,\\
\rho_0, & r_h<r<r_{\rm heat},\\
\rho_0\times\frac{\rho_{\rm AC}(r)}{\rho_{\rm AC}(r_{\rm heat})},
& r>r_{\rm heat}.
\end{array}\right.
\end{equation}
The inner most density profile ($\propto r^{-3/2}$) corresponds to
the collisionally regenerated structures (``crest'') of DM due to
the joint evolution of baryons and DM in the enviroment of a central
black hole \cite{2007PhRvD..75d3517M}. For NGC 6388, $M_{\rm IMBH}\sim
6\times 10^3$ M$_{\odot}$, $r_h\sim 0.4$ pc is found with the
observational baryon density. For M 15 there is no consensus of the
existence of IMBH \cite{1996A&A...315..396D,2002AJ....124.3270G}, and
we employ Eq. (2.3) to describe the DM density profile of M 15 without
considering the possible IMBH.

\subsection{Astrophysical $J$-factor}

For Majorana fermion DM particles, the $\gamma$-ray flux from DM 
annihilation can be written as
\begin{equation}
\label{eqn:phi}
\Phi(\Delta\Omega,E_{\gamma})=\frac{1}{4\pi} \times \frac{\langle
\sigma v\rangle}{2m^2_{\chi}}\,\frac{\mathrm{d}N_{\gamma}}{\mathrm{d}
E_{\gamma}} \times \bar{J}(\Delta\Omega)\Delta\Omega,
\end{equation}
where $m_{\chi}$ and $\langle\sigma v\rangle$ are the mass and velocity
weighted thermal average annihilation cross section of DM particles,
${{\rm d}N_{\gamma}/{\rm d}E_{\gamma}}$ is the $\gamma$-ray
spectrum for one annihilation. The astrophysical factor ($\bar{J}$) is
the integral of the density square along the line of sight (LOS) averaged
over the solid angle $\Delta\Omega$
\begin{equation}
\label{eqn:jbar}
\bar{J}(\Delta\Omega) = \frac{1}{\Delta\Omega}
\int_{\Delta\Omega}{\rm d}\Omega\int_{\rm LOS}{\rm d}l\,\rho^2(r(l)).
\end{equation}

For H.E.S.S. observations, the integral solid angle is $\Delta\Omega=
5\times 10^{-6}$ sr, which corresponds to a cone with half angle
$0.07^{\circ}$ \cite{2011ApJ...735...12A}. Since the resolution angle
of {\it Fermi}-LAT in GeV range is much larger ($>0.5^{\circ}$,
\cite{2009ApJ...697.1071A}), we need to enlarge the integral solid
angle. The tidal radii of both GCs are about $30$ pc, and the distances
are about $10$ kpc \cite{2011ApJ...735...12A}. The opening angles of
these two GCs are $\sim 0.17^{\circ}$. Therefore they can be regarded
as point sources for {\it Fermi}-LAT and we integrate all the DM
contribution to the tidal radius to calculate the $J$-factor.
It is found that the final $J\times\Delta\Omega$ is about $7.8\times
10^{19}$ ($3.4\times10^{20}$) GeV$^2$ cm$^{-3}$ for M 15 (NGC 6388),
which is larger by $\sim10\%$ ($0.1\%$) compared with that within
$0.07^{\circ}$ cone as adopted by H.E.S.S.. The $J$-factor of NGC 6388 is
larger than M 15 is mainly due to the difference of the density profiles
after AC which depends on the final baryon density profiles of the GCs,
and the heating effect. Compared with the dwarf galaxies as given in
\cite{2011PhRvL.107x1302A}, the $J$-factors of GCs are generally larger,
which is also due to the AC process.

\section{{\it Fermi}-LAT data analysis}

The {\it Fermi}-LAT data\footnote{http://fermi.gsfc.nasa.gov/ssc/data}
used in this analysis are the new ``Pass 7'' data recorded between 4
August 2008 and 2 September 2011. Photons with Event Class ``Source''
(evclass=2) and zenith angle within $100^{\circ}$
are selected. The energy range of events is cut from $200$ MeV to
$300$ GeV, and the radius of region-of-interest (ROI) is adopted
to be $6^{\circ}$. We use the LAT Scientific Tools v9r23p1 to do this
analysis. The unbinned likelihood analysis method is adopted. The
instrument response function used is ``{\tt P7SOURCE\_V6}''.
For the diffuse background, we use the Galactic diffuse model
{\tt gal\_2yearp7v6\_v0.fits} and the isotropic background spectrum
{\tt iso\_p7v6source.txt} provided by the {\it Fermi} Science Support
Center\footnote{http://fermi.gsfc.nasa.gov/ssc/data/access/lat/BackgroundModels.html}.

\subsection{NGC 6338}

The detection of $\gamma$-ray emission from NGC 6388 was reported
in \cite{2010A&A...524A..75A}, with {\it Test Statistic}
\cite{1996ApJ...461..396M} value TS$=86.6$. The spectral energy
distribution can be fitted using power-law function with an exponential
cutoff $E^{-\Gamma}\exp(-E/E_c)$, which is expected for the emission from
a population of millisecond pulsars (MSP). The best-fitting parameters
are $\Gamma=1.1^{+0.7}_{-0.5}$ and $E_c=1.8^{+1.2}_{-0.7}$ GeV
\cite{2010A&A...524A..75A}. Using the likelihood tool {\tt gtlike} we
re-do the spectral analysis with more data. The source model XML file
is generated using the user contributed tool {\tt
make2FGLxml.py}\footnote{http://fermi.gsfc.nasa.gov/ssc/data/analysis/user/}
based on the 2FGL source catalog \cite{2011arXiv1108.1435T}. The spectrum
of NGC 6388 is also modelled with $E^{-\Gamma}\exp(-E/E_c)$. By setting
all the source parameters within the ROI free, the best-fitting parameters
for NGC 6388 are $\Gamma=1.21\pm0.17$ and $E_c=1.82\pm0.35$ GeV, with
a TS value $596$. The fitting parameters are consistent with that given 
in \cite{2010A&A...524A..75A}.

To derive the spectral energy distribution (SED) of NGC 6388, we divide
the data into different energy bins, and use {\tt gtlike} tool to fit
the parameters for each bin. Two methods are adopted in the fit. We
first fix the parameters of all other sources and the normalizations
of diffuse backgrounds derived above in the global fit, leaving only
the normalization parameters of NGC 6388 and the very bright pulsar
PSR J1709-4429 free. The spectral parameters of NGC 6388 and PSR
J1709-4429 are also fixed to be the best fitting values. Because the
energy bin is relatively narrow the precise values of the spectral
parameters have little effect on the final results
\cite{2010A&A...524A..75A}. The results are shown by the filled
circles in Figure \ref{fig:spectrum}. The solid line shows the
fitting curve with spectrum $E^{-\Gamma}\exp(-E/E_c)$, as a represent
of MSP-type emission. It shows good agreement between the global fit
and the individual fit for different energy bins. Then we relax the
normalization parameters of all the sources and the diffuse backgrounds,
and re-do the fit. The results are shown by the empty circles in Figure
\ref{fig:spectrum}. We can see that there are some differences between
the results of these two methods, which might be originated from the
complexity of the diffuse background models.

\FIGURE{
\includegraphics[width=0.6\columnwidth]{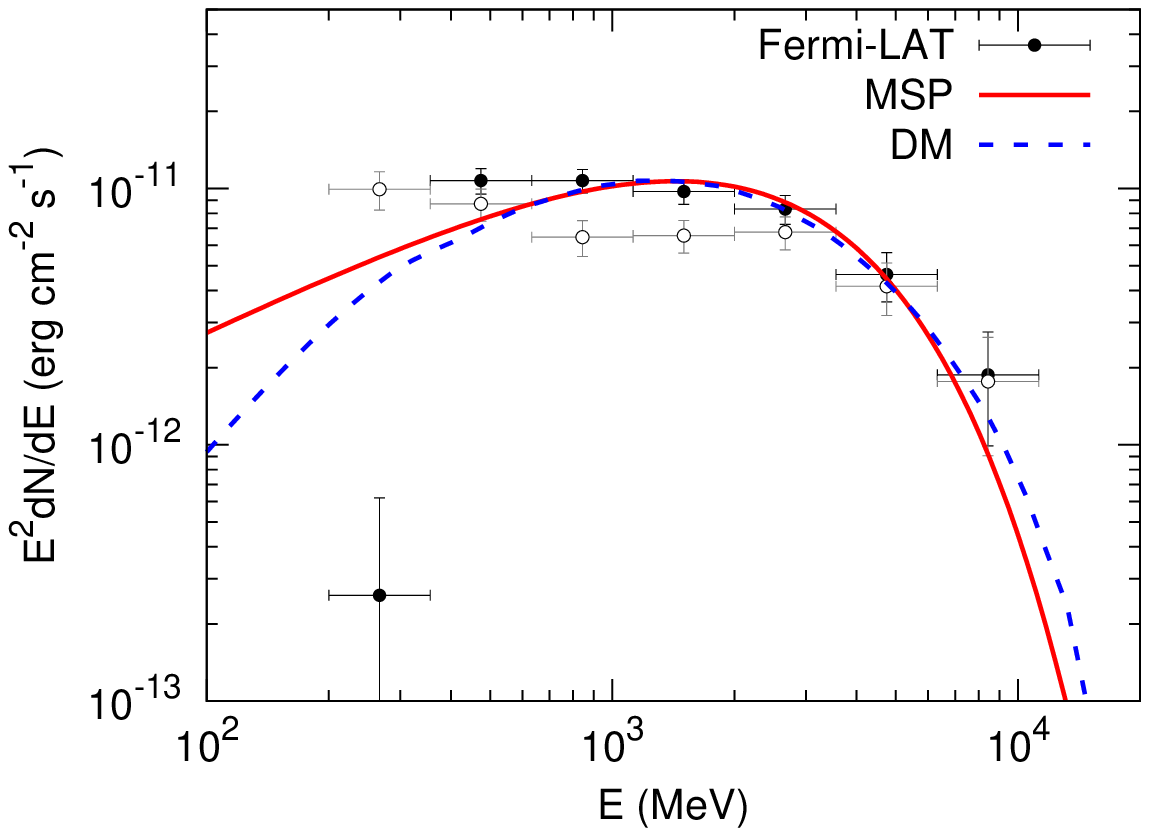}
\caption{SED of NGC 6388.}
\label{fig:spectrum}
}

We then add a DM component at the position of NGC 6388 to search for
possible DM contribution to the emission. Different annihilation final
states are investigated, including $b\bar{b}$, $W^+W^-$, $\mu^+\mu^-$,
$\tau^+\tau^-$ and $\gamma$-ray line. We use PYTHIA simulation tool to
calculate the $\gamma$-ray yield spectrum for each final state
\cite{2006JHEP...05..026S}. A series values of DM mass from $10$ GeV to
$10$ TeV are considered. For monochromatic $\gamma$-ray line we use
Gaussian function to model its spectral shape. The energies of photons
from $300$ MeV to $200$ GeV are searched. The width of Gaussian function
as a function of photon energy is adopted to be the energy resolution of
{\it Fermi}-LAT, which is about $8\%-13\%$ ($\Delta E/E$) in this energy
range \cite{2009ApJ...697.1071A}. Given the spectrum shape of DM
contribution, we use the python likelihood
tool\footnote{http://fermi.gsfc.nasa.gov/ssc/data/analysis/scitools/python\_tutorial.html}
to fit the normalization parameter and derive the flux upper limits.

It is found that the detected $\gamma$-ray spectrum of NGC 6388 can also
be fitted with a DM component, with mass $\sim 25$ GeV and annihilation
final state $b\bar{b}$, as shown by the dashed line in Figure
\ref{fig:spectrum}. Note that the recent analysis of the {\it Fermi}-LAT
data in the Galactic center region also showed possible additional
emission compatible with DM contribution with $m_{\chi}\sim30$ GeV
for $b\bar{b}$ annihilation final state \cite{2011arXiv1110.0006H}.

It is well motivated that the $\gamma$-ray emission from GCs may come
from the MSPs, therefore we do not claim the DM origin of the $\gamma$-rays
of NGC 6388. In any case we can instead set an upper limit of the
contribution from DM annihilation. Here the upper limits are derived
for different final states and different mass of DM particle individually.
We use two ways to fit the data. The first one (Fit 1) is to fix
all the parameters of sources derived in the above global fit. The free
parameters are the normalizations of the diffuse backgrounds and the DM
component. The other way (Fit 2) is to leave the normalizations of sources
in the ROI and the diffuse backgrounds as well as the normalization
of the DM component free. The spectral parameters of the sources are
fixed to be the best fitting values in the global fit. Fit 1 corresponds
to a more stringent constraint, based on the assumption that the observed
$\gamma$-rays come from astrophysical sources. Fit 2 gives a weaker but
more conservative constraint. The $95\%$ confidence level (C.L.) upper
limits of the $>100$ MeV fluxes from DM annihilation for NGC 6388
are shown in the left panel of Figure \ref{fig:ul}. The thick and
thin lines correspond to Fit 1 and 2 respectively. The same way is
applied to the line analysis. The upper limits of line emission are
shown in Figure \ref{fig:ulline}.

\FIGURE{
\includegraphics[width=0.45\columnwidth]{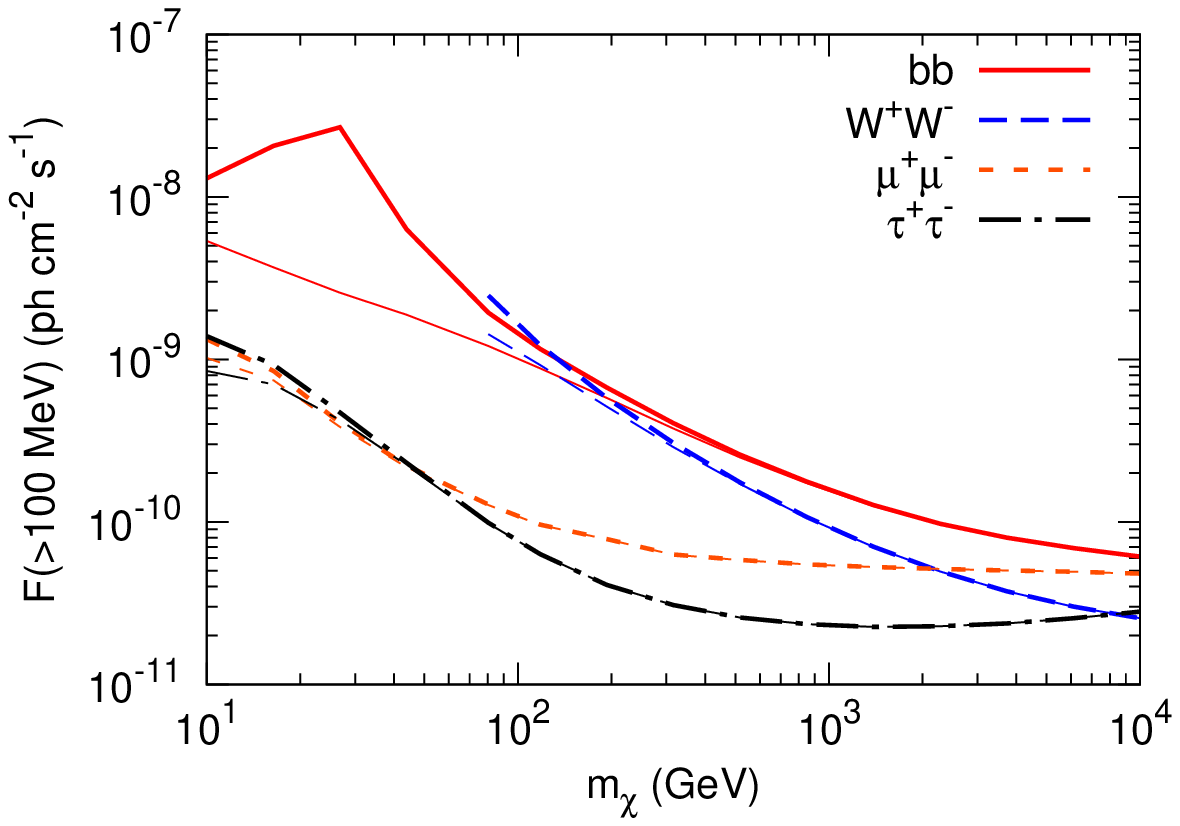}
\includegraphics[width=0.45\columnwidth]{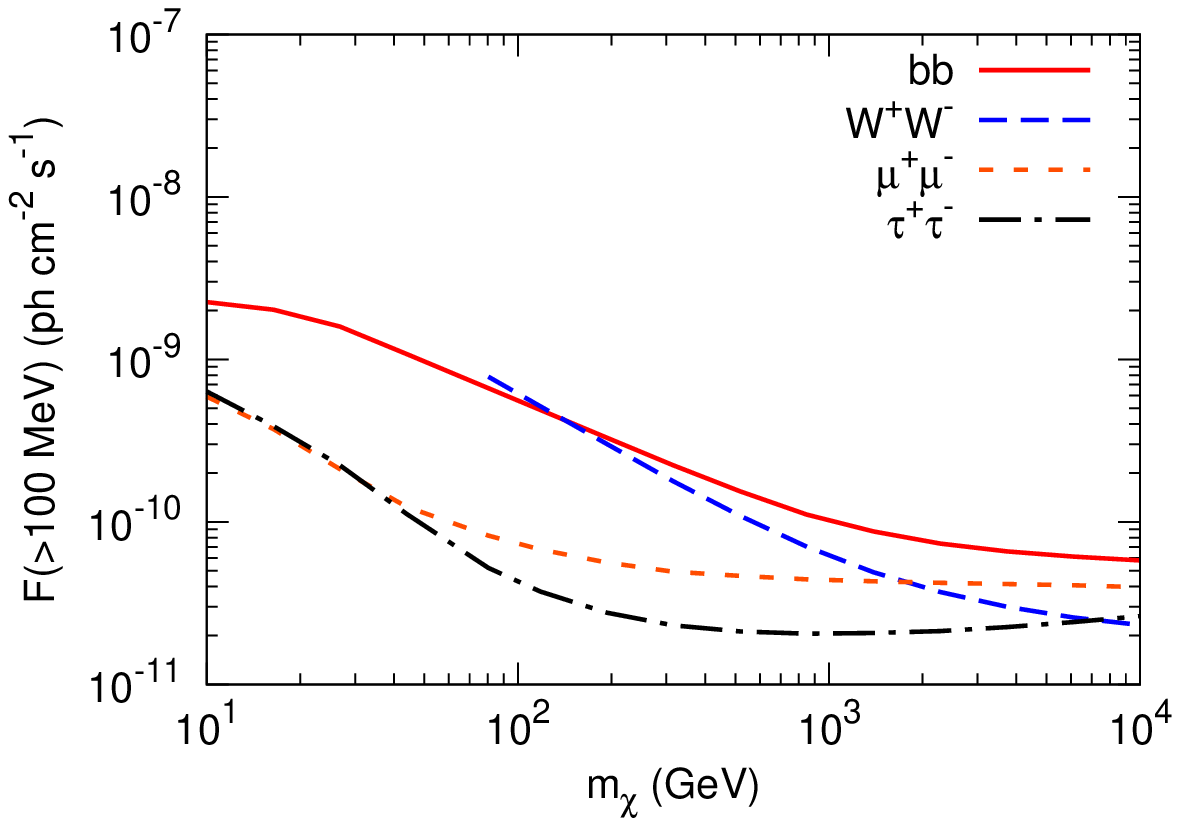}
\caption{Derived 95\% C.L. upper limits of the WIMP annihilation
contribution to the $>100$ MeV $\gamma$-ray fluxes as functions of WIMP
mass $m_{\chi}$, for GCs NGC 6388 (left) and M 15 (right).}
\label{fig:ul}
}

\FIGURE{
\includegraphics[width=0.6\columnwidth]{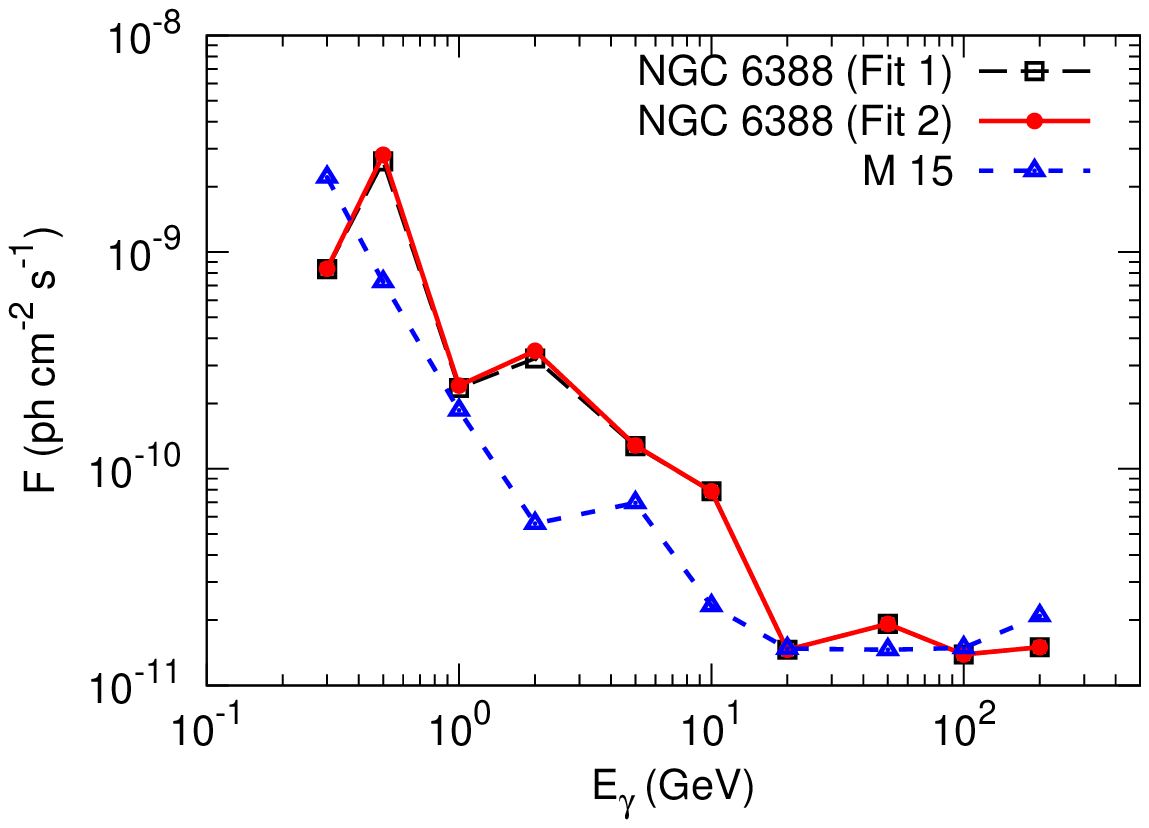}
\caption{95\% C.L. upper limits of the monochromatic $\gamma$-ray
emission as functions of $\gamma$-ray energy.}
\label{fig:ulline}
}

\subsection{M 15}

There was no firm detection of emission from M 15 in the previous
analysis. In \cite{2010A&A...524A..75A} M 15 was reported to have
a very weak signal with TS$=5.4$. In our analysis with more data, we
find a relatively higher TS value $\sim12$, for both power-law
and power-law + exponential cutoff models. We can derive the upper
limits on DM annihilation to $\gamma$-rays, similar as the
analysis of NGC 6388. Because there is no detection of $\gamma$-rays,
only the method Fit 2 is adopted for M 15.
There are three other sources in the ROI of M 15, 2FGL J2115.4+1213,
2FGL J2112.5+0818 and 2FGL J2147.3+0930. The free parameters include
the spectral parameters of M 15 (power-law model for a possible MSP
contribution) and these three sources, the normalizations of the
diffuse backgrounds and the DM component of M 15. The results are
shown in Figures \ref{fig:ul} and \ref{fig:ulline}.

\section{Constraints on DM models}

Integrating Eq. (\ref{eqn:phi}) above $100$ MeV we can easily get the
upper limits of the DM annihilation cross section using the flux upper
limits. The results are presented in Figures \ref{fig:svbw} -
\ref{fig:svline} respectively.

Figure \ref{fig:svbw} shows the constraints on $m_{\chi}-
\langle\sigma v\rangle$ for $b\bar{b}$ (left) and $W^+W^-$ (right)
final states. For comparison we also show the
results from the combined analysis of {\it Fermi} observations of $10$
dwarf galaxies \cite{2011PhRvL.107x1302A} and that given by H.E.S.S.
observation of these two GCs ($W^+W^-$, \cite{2011ApJ...735...12A}).
It is shown that the constrains given in this work are generally
stronger than that for dwarf galaxies, at least for DM mass $\gtrsim50$
GeV. Because of the high density spike from the AC process, the
distribution of DM is more concentrated than the initial NFW profile.
Although the heating effect from stars will smooth out the central density
spike, the $J$-factors of GCs are still higher than that of dwarf galaxies
(e.g., estimated with NFW profiles), and the constraints on DM models are
stronger accordingly.
For $b\bar{b}$ final state, the DM annihilation induced $\gamma$-ray
spectrum is similar with the observed data of NGC 6388, therefore the
constraint is a bit weaker when $m_{\chi}<50$ GeV. However, for the
method Fit 1 the constraint is always stronger than that for dwarf
galaxies. H.E.S.S. constraints are more effective for massive DM
($m_{\chi}\gtrsim2$ TeV).

Also shown in Figure \ref{fig:svbw} are the theoretically expected
neutralino annihilation cross sections (multiplied with the branching
ratios to $b\bar{b}$ and $WW+ZZ$) in the Minimum Supersymmetric
Standard Model (MSSM). In the right panel we sum the model predicted
cross sections to $W^+W^-$ and $ZZ$ chanels together due to the similarity
of $\gamma$-ray spectra from these two channels. We utilize numerical code
micrOMEGAs \cite{2007CoPhC.176..367B,2009CoPhC.180..747B} to perform a
random scan in the 7-dimensional parameter space at the electroweak scale.
These parameters include the CP-odd Higgs mass $m_A$, the Higgs mixing
mass parameter $\mu$, the wino mass parameter $M_2$, the sfermion mass
parameter $m_{\tilde{f}}$, the ratio of two Higgs vacuum expectation
values $\tan \beta$, the trilinear parameters of the third family
squark $A_{\tilde{b}}$ and $A_{\tilde{t}}$. The other trilinear
parameters are set to zero. We also impose the assumptions that the
gaugino mass parameters are related by $M_1:M_2:M_3=\alpha_1:\alpha_2:
\alpha_3$ for grand unification, where the $\alpha_i$ are the coupling
constants of three standard model gauge groups, and $M_1:M_2=
\frac{5}{3} \tan^2 \theta_{_W}$. The ranges of the parameters are taken as
follows: 50 GeV$< |\mu|, M_2 <$10 TeV, 100 GeV $<m_A, m_{\tilde{f}}<$1 TeV,
 1$<\tan \beta<$60, $-5m_{\tilde{f}} <A_t, A_b<
5m_{\tilde{f}}$ and ${\rm sign}(\mu)=\pm 1$. Several constraints from
accelerator experiments and DM detection are implemented in our
numerical scan. We set the limit for $\rho$ parameter as $\rho-1<2.2
\times 10^{-3}$\cite{2006JHEP...33..0603}. Some important flavor physics
constraints include: $Br(B \to X_s \gamma) = (3.55 \pm 0.24) \times
10^{-4}$ \cite{bsgamma}, $Br(B_s \to \mu^+ \mu^- ) = (0 \pm 1.4)
\times 10^{-8}$ \cite{2010PhLB..693..539D,2011PhLB..699..330L},
$Br(B_u \to \tau \nu )/Br(B_u \to \tau \nu )_{\rm SM} = 1.28 \pm 0.38$
\cite{bsgamma}. Here we only require the supersymmetric contributions
to satisfy these constraints at $3\sigma$ level, and adopt a very
conservative bound for muon anomalous magnetic moment
\cite{2011EPJC...71.1515D} as $-11.4 \times 10^{-10} < \delta \alpha_\mu
< 9.4 \times 10^{-9}$ \cite{2006JHEP...33..0603}. The mass bound of
standard model like Higgs $m_h> 114$ GeV, limits on the masses of
light charge sparticle from LEP (for details, see \cite{2007CoPhC.176..367B,
2009CoPhC.180..747B}), and DM direct detection constrains from XENON100
\cite{2011PhRvL.107m1302A} are also taken into account.

In Figure \ref{fig:svbw} the squares are for DM models which can give
the right relic density \cite{2011ApJS..192...18K} if DM is thermally
produced in the early Universe (labelled as ``MSSM-thermal''). The
triangles are the cases with thermal relic density not higher than the
measured value, the correct DM abundance in these models could be
produced via some non-thermal mechanisms. We can see that for $b\bar{b}$
final state the current constraint can reach the natural scale for
thermally produced DM with mass of $O(10)$ GeV. In the MSSM model,
large neutralino annihilation cross section to $b\bar{b}$ final states
could arise from the resonance effect with $m_{A}\sim 2m_{DM}$. Some
non-thermal models with this feature have been excluded by our
constraints. For $W^+W^-$ channel the constraint is a bit weaker but also
close to the natural scale. In the MSSM scenario, large higgsino or wino
component (depending on the relations between the three parameters $M_1$,
$M_2$ and $\mu$) in the neutralino would enhance DM annihilation cross
section to gauge bosons significantly. If the neutralino mass lies in the
range $(80,\,300)$ GeV, many non-thermal models would be also stringently
constrained by our results.

\FIGURE{
\includegraphics[width=0.45\columnwidth]{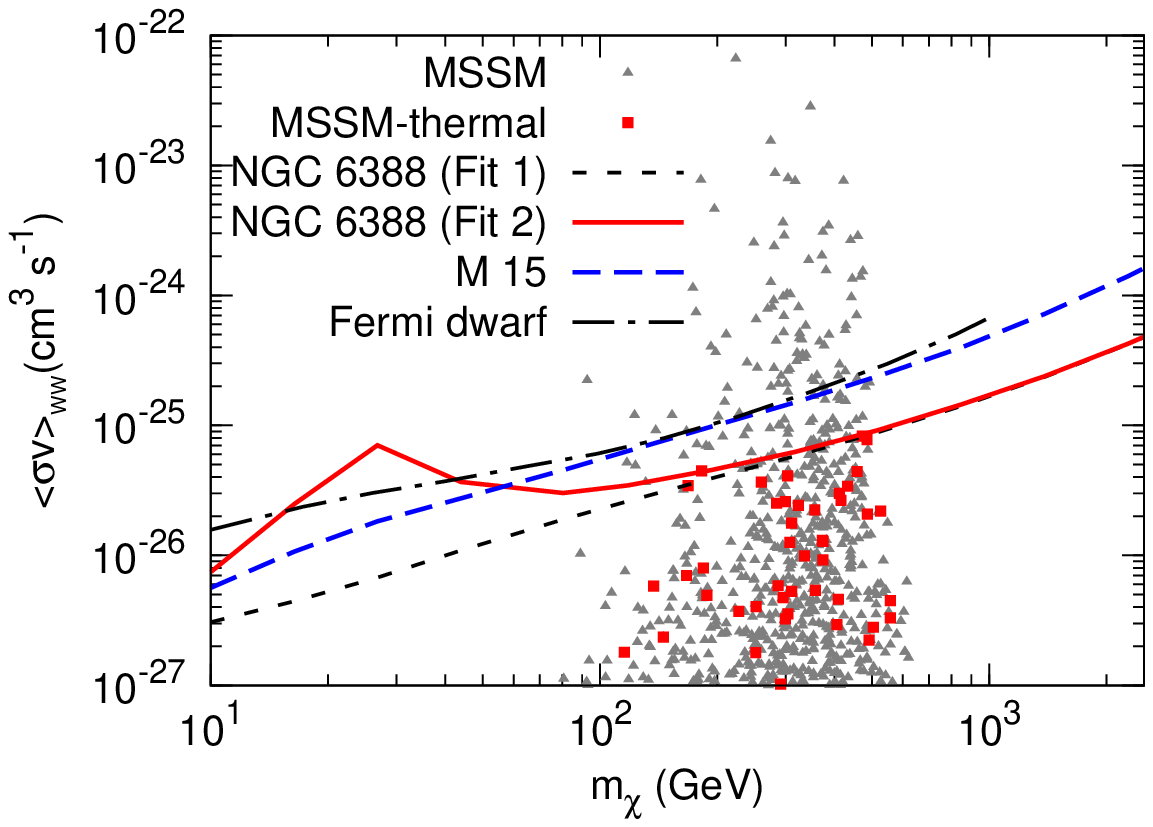}
\includegraphics[width=0.45\columnwidth]{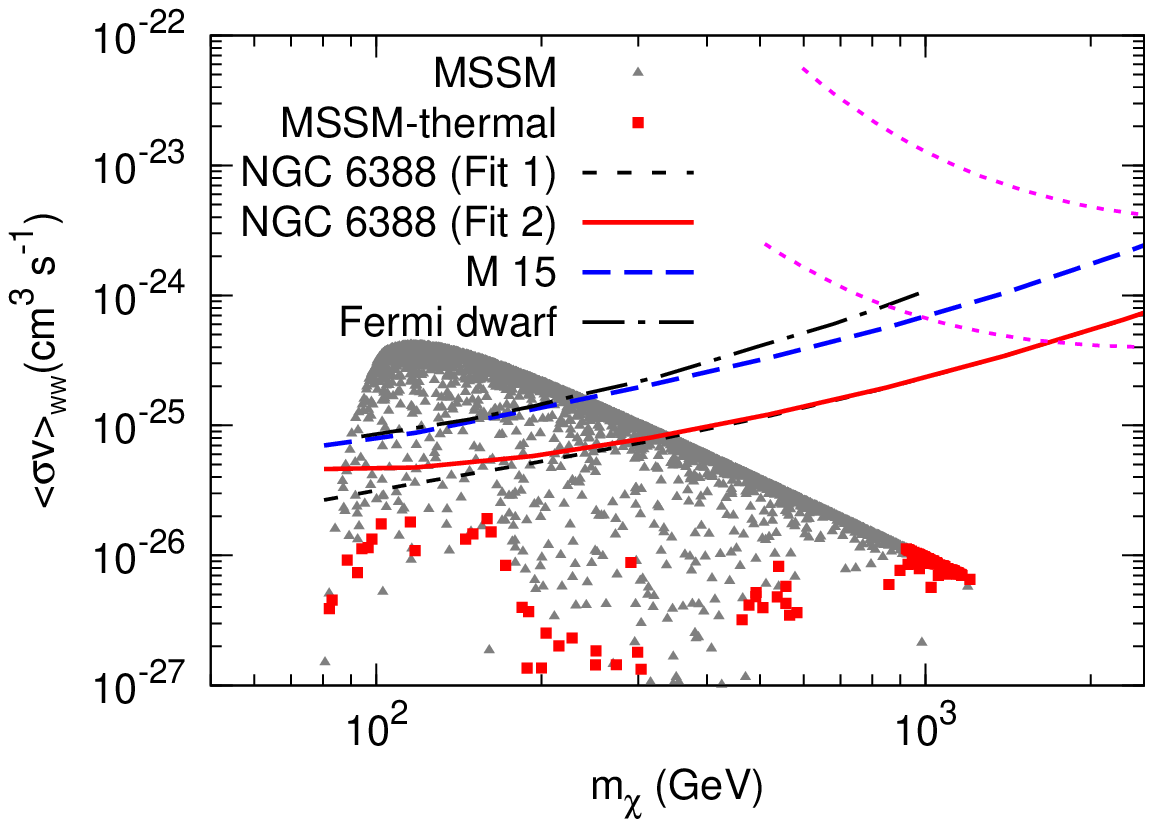}
\caption{Constraints on DM mass vs. annihilation cross section, for
$b\bar{b}$ (left) and $W^+W^-$ (right) final states. Points are a random
scan of the MSSM parameter space taking into account the current
constraints from accelerator data. Magenta dotted lines in the right
panel are the constraints got by H.E.S.S. observations of NGC 6388
(lower) and M 15 (upper).}
\label{fig:svbw}
}

\FIGURE{
\includegraphics[width=0.45\columnwidth]{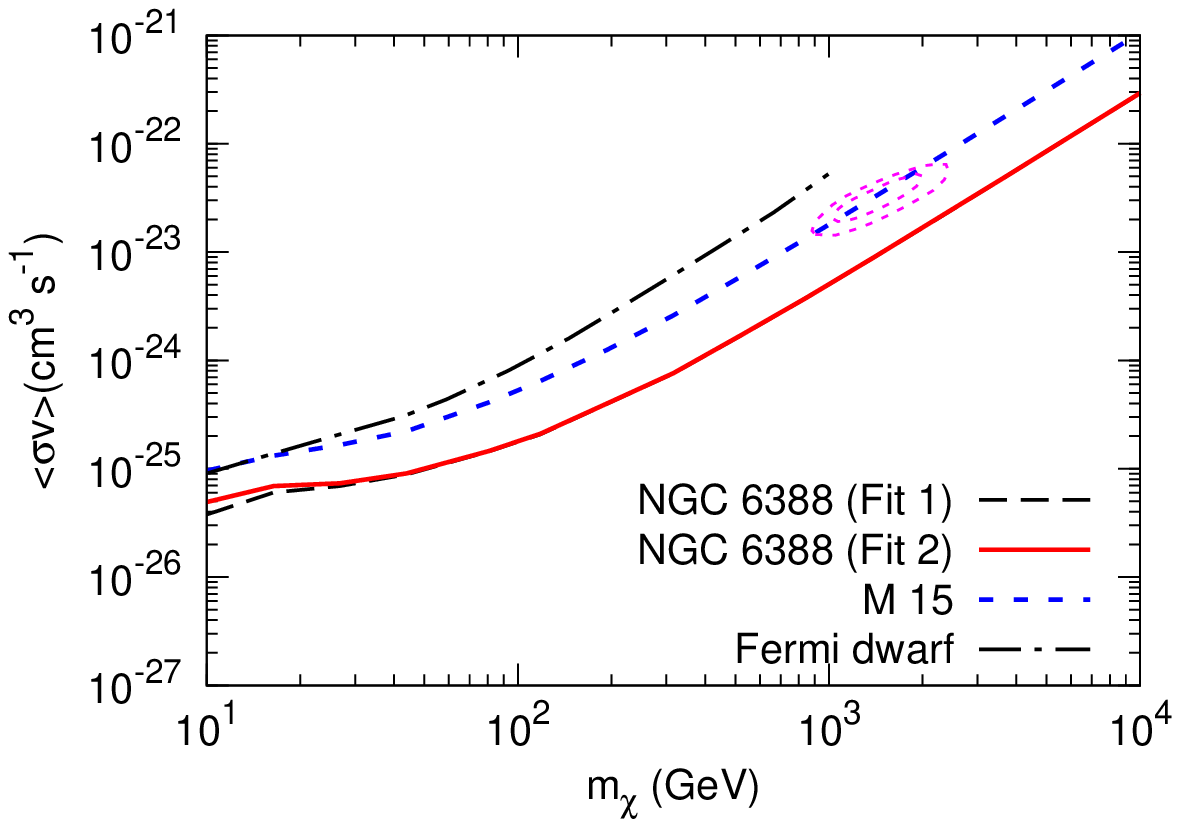}
\includegraphics[width=0.45\columnwidth]{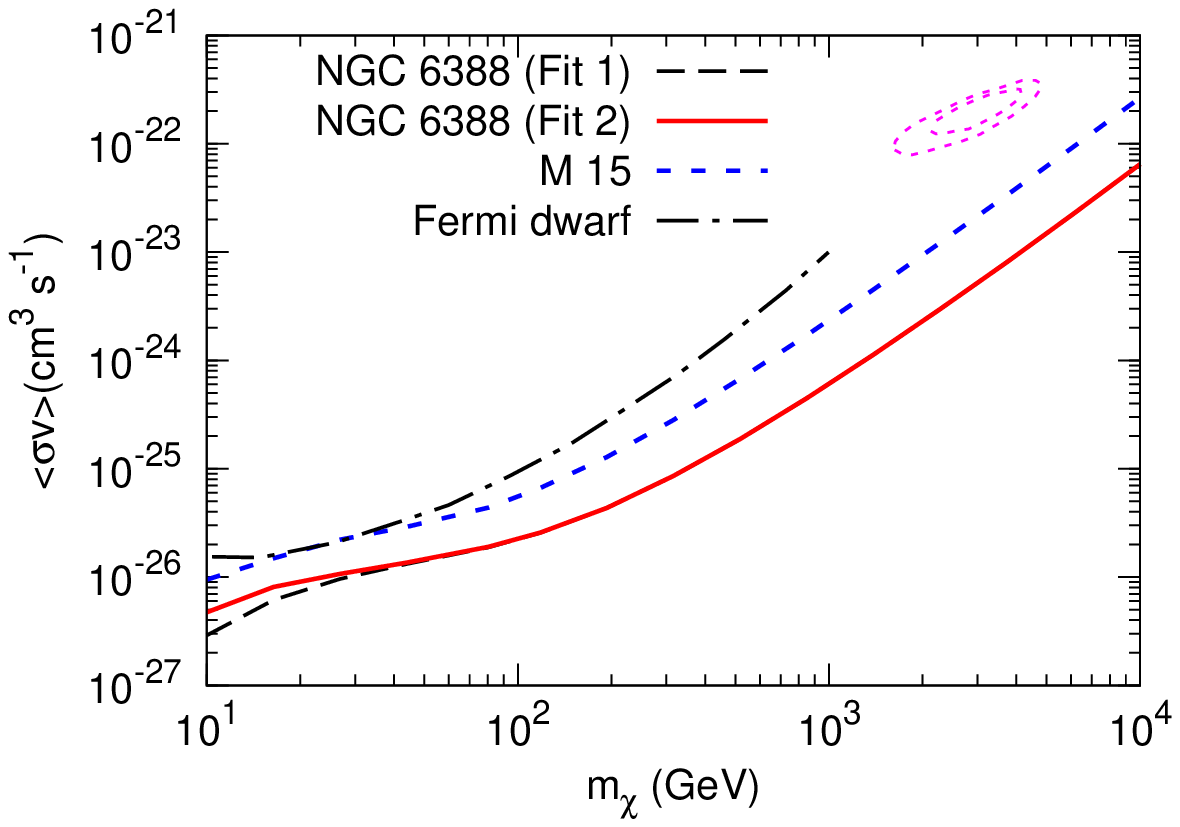}
\caption{Same as Fig. \ref{fig:svbw} but for leptonic final states
$\mu^+\mu^-$ (left) and $\tau^+\tau^-$ (right) respectively.}
\label{fig:svmt}
}

\FIGURE{
\includegraphics[width=0.45\columnwidth]{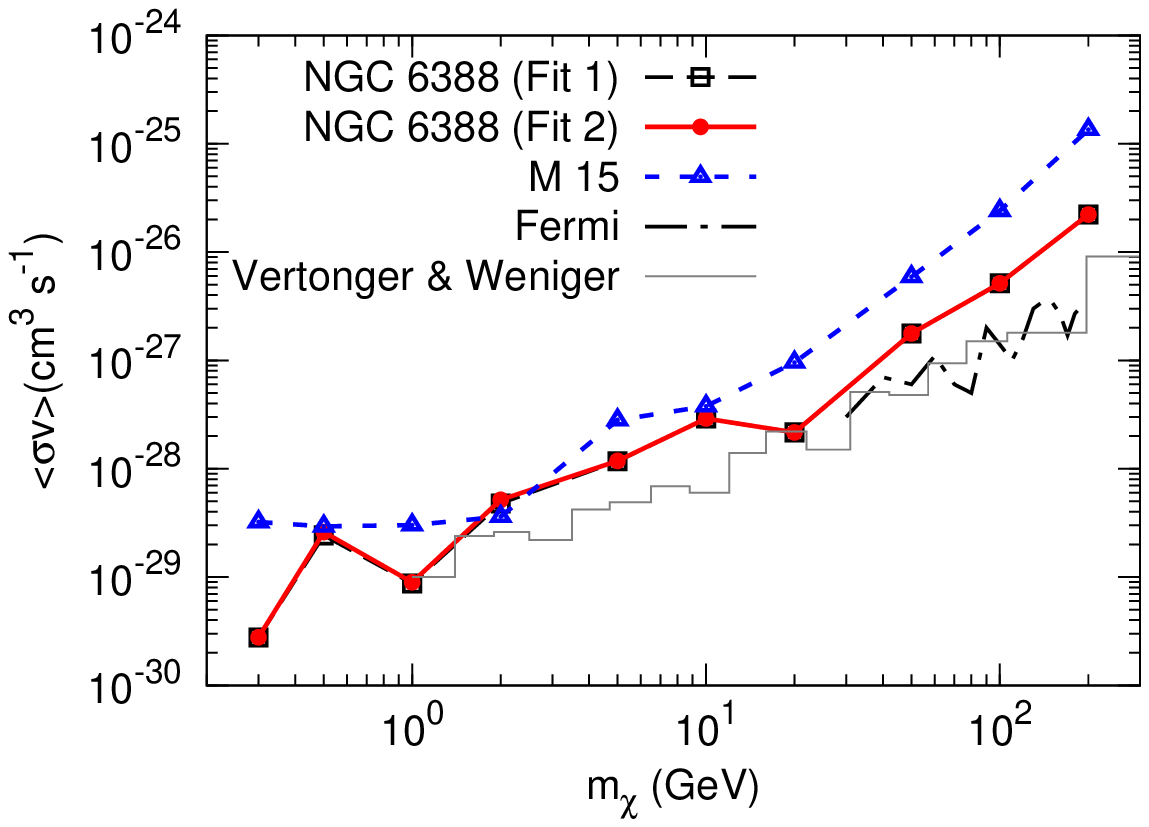}
\includegraphics[width=0.45\columnwidth]{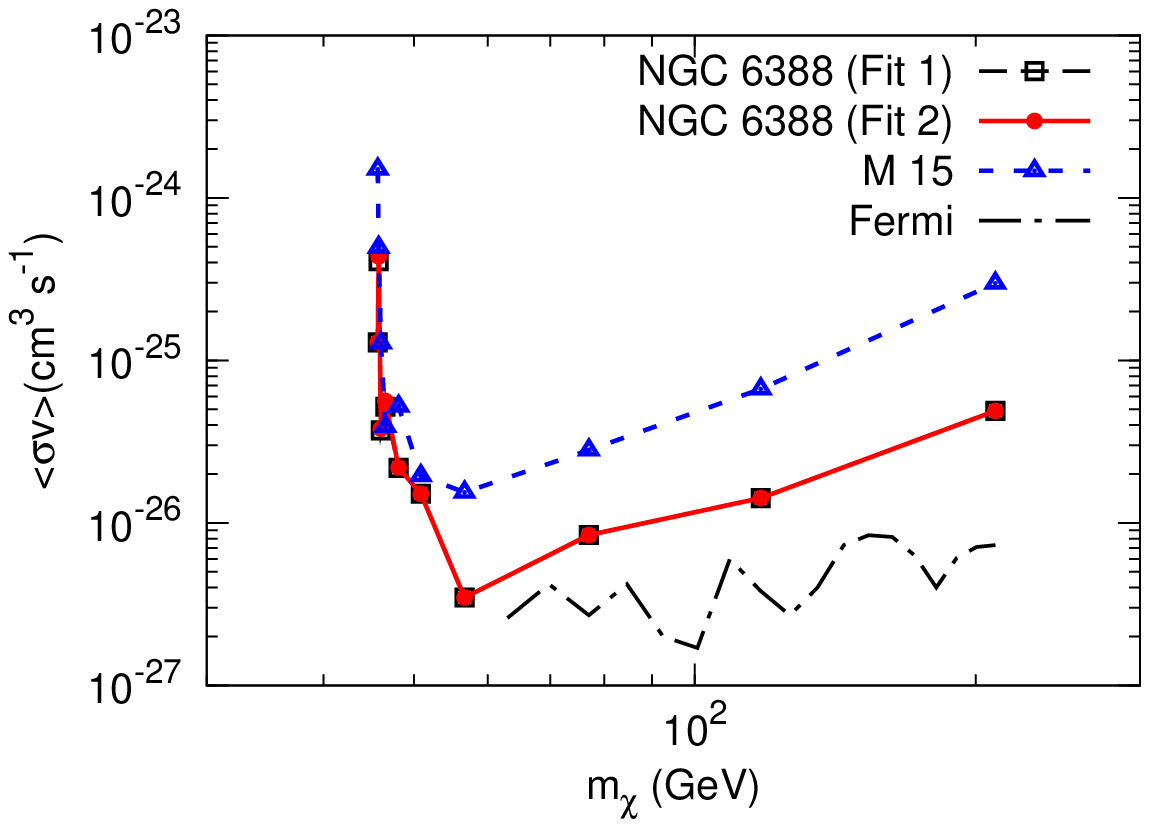}
\caption{Same as Fig. \ref{fig:svbw} but for DM annihilation into
$\gamma\gamma$ (left) and $\gamma Z^0$ (right) final states.}
\label{fig:svline}
}

Motivated by the recent observations of the CR positron/electron excesses
at PAMELA, ATIC and Fermi-LAT
\cite{2009Natur.458..607A,2008Natur.456..362C,2008PhRvL.101z1104A,
2009A&A...508..561A,2009PhRvL.102r1101A}, and the non-excess of antiprotons
\cite{2009PhRvL.102e1101A,2010PhRvL.105l1101A}, the leptonic DM models
are proposed to explain the data (e.g., \cite{2009NuPhB.813....1C,
2009PhRvD..79b3512Y,2009PhRvL.103c1103B,2010NuPhB.831..178M,
2010PhRvD..81b3516L,2012PhRvD..85d3507L}). We also study the constraints
on the leptonic final states $\mu^+\mu^-$ and $\tau^+\tau^-$ by DM
annihilation which might
be responsible for the $e^{\pm}$ excesses. Here the inverse Compton
scattering $\gamma$-rays generated by the decaying products $e^{\pm}$
from muons or tauons are not taken into account. As illustrated in
\cite{2010ApJ...712..147A}, considering the inverse Compton $\gamma$-rays
the constraints on DM model parameters will be stronger, however,
depending on the uncertainties of the diffusion process of $e^{\pm}$. Thus
the results given here should be conservative. Constraints on $m_{\chi}-
\langle\sigma v\rangle$ parameter space are shown in Figure \ref{fig:svmt}.
The contours show the favored parameter region to fit the CR $e^{\pm}$
data \cite{2010NuPhB.831..178M}. It is shown that the models to explain
the $e^{\pm}$ excesses should be excluded by the {\it Fermi}-LAT data
about GCs.

Finally we study the constraints on possible monochromatic $\gamma$-ray
line emission from e.g., $\chi\chi\to\gamma\gamma$ or $\chi\chi\to\gamma
Z$ of DM annihilation. No significant line emission is found in the
data. The upper limits of line emission are derived. The constraints on
cross sections to $\gamma\gamma$ and $\gamma Z$ are given in Figure
\ref{fig:svline}. Note we have $E_{\gamma}=m_{\chi}$ for
$\chi\chi\to\gamma\gamma$, and $E_{\gamma}=m_{\chi}(1-m_Z^2/4m_{\chi}^2)$
for $\chi\chi\to\gamma Z$. The results derived with {\it Fermi}-LAT data
including the Galactic center region by {\it Fermi} collaboration
(NFW profile, \cite{2010PhRvL.104i1302A}) and Vertonger \& Weniger
\cite{2011JCAP...05..027V} are also shown for comparison. Our
constraints are a bit weaker than the results in these two works.
We think it is reasonable because their analysis regions are much
larger and include the Galactic center region, which will give a higher
$J$-factor of DM annihilation.

\section{Conclusions and Discussions}

The GCs are thought to form in the cosmological context with AC
process at the beginning which pulls DM into the halo center and
results in a very high annihilation luminosity. Thus search for
$\gamma$-rays from GCs may be effective to probe the particle nature
of DM. In this work we analyze the {\it Fermi}-LAT three-year data
(Pass 7) of GCs NGC 6638 and M 15 and constrain the DM annihilation
models. A clear detection of $\gamma$-ray emission from NGC 6388 is
found, with TS value $\sim600$. The spectrum of NGC 6388 can be well
fitted with a power-law + exponential cutoff function, which is
expected for the emission of a population of MSPs. We find that
a DM scenario with $m_{\chi}\sim25$ GeV and $b\bar{b}$ final state
can also fit the SED. For M 15 no significant $\gamma$-ray emission
if found (the spectral fit indicates a potential source with
TS $\approx12$).

Assuming there is an additional spectral component from DM annihilation
of these two GCs, we derive the upper limits of the DM component for
different DM masses ($10\,{\rm GeV}-10\,{\rm TeV}$) and annihilation
final states ($b\bar{b},\,W^+W^-,\,\mu^+\mu^-,\,\tau^+\tau^-,\,
\gamma\gamma,\,\gamma Z$) (Figures \ref{fig:ul} and \ref{fig:ulline}).
The constraints on the DM annihilation cross section are given
(Figures \ref{fig:svbw}-\ref{fig:svline}). Except for the line
emissions, the constraints are stronger than that derived
according to the {\it Fermi}-LAT observations of dwarf galaxies. For
DM mass smaller than TeV our constraints are also stronger than that
given by H.E.S.S. observations of the same GCs. For $b\bar{b}$ and
$W^+W^-$ final states which are generally expected from supersymmetric
DM model, the constraints can reach the natural scale with which DM
is thermally produced. Especially the leptonic annihilation models
to explain the CR $e^{\pm}$ excesses can be excluded by the current
analysis.

However, the uncertainties of the present analysis, for example the
properties of the hypothetical DM halo and the origin and evolution
of the GCs, are far from clear. The GCs were assumed to be formed in
the cosmological context, and the DM density profiles in the GCs are
modelled taking into account the most probable astrophysical processes,
e.g., the AC by baryons, adiabatic growth of an IMBH and the scattering
by stars. Future studies on the observations and modelings of the DM
distribution in the GCs are necessary to improve the current work.

\acknowledgments

We thank Yi-Zhong Fan, Rui-Zhi Yang and Xiao-Yuan Huang for discussion.
This work is supported by the Natural Science Foundation of China under
the grant Nos. 11075169, 11075074, 11065004, 11105155, 11105157 and
11175251, the 973 project under grant No. 2010CB833000 and the Chinese
Academy of Science under Grant No. KJCX2-EW-W01. L. Feng is supported
by the Research Fund for the Doctoral Program of Higher Education under
grant No. 200802840009.

\bibliographystyle{JHEP}
\bibliography{/home/yuanq/work/cygnus/tex/refs}

\end{document}